\documentclass[doublespacing]{elsart}
\usepackage[english]{babel}

\usepackage{epsfig}
\usepackage{graphics}

\usepackage{amssymb}

\begin{document}

\begin{frontmatter}

\title{Growth of Interaction Between Antiprotons (Negative Hyperons) and a Nuclear Pseudomagnteic Field Under Deceleration in
Matter with Polarized Nuclei: The Possibility  to Study the
Spin--Dependent Part of the Forward Scattering Amplitude  in the
 Range of Low--Energies}

\author{V.G. Baryshevsky}
\begin{abstract}
The influence of Coulomb interaction on the phenomenon of
"optical" spin rotation of negatively charged particles
(antiprotons, etc.) moving in matter with polarized nuclei is
considered. It is shown that because the density of the antiproton
(negative hyperon) wave function on the nucleus increases, the
spin precession frequency grows as the particle decelerates. As a
result, spin rotation of negatively charged particles becomes
observable despite their rapid deceleration. This provides
information about the spin--dependent part of the amplitude of
coherent elastic zero--angle scattering in the range of low
energies, where scattering experiments are practically impossible
to perform.
\end{abstract}


%
\begin{abstract}
The influence of Coulomb interaction on the phenomenon of
"optical" spin rotation of negatively charged particles
(antiprotons, etc.) moving in matter with polarized nuclei is
considered. It is shown that because the density of the antiproton
(negative hyperon) wave function on the nucleus increases, the
spin precession frequency grows as the particle decelerates. As a
result, spin rotation of negatively charged particles becomes
observable despite their rapid deceleration. This provides
information about the spin--dependent part of the amplitude of
coherent elastic zero--angle scattering in the range of low
energies, where scattering experiments are practically impossible
to perform.
\end{abstract}
\end{frontmatter}

\section*{Introduction}

The advent of the Facility for Low-Energy Antiproton and Ion
Research has spurred the rapid development of low--energy
antiproton physics (FLAIR) \cite{1,2}.

The possibility to obtain  polarized antiprotons by a
spin-filtering method \cite{3} opens up opportunities for
investigation of a large number of spin--dependent fundamental
processes arising when antiprotons pass through matter with
polarized nuclei (protons, deuterons, $^3$He, etc.) In particular,
study of the phenomenon of particle "optical" spin rotation in a
nuclear pseudomagnetic field of matter with polarized nuclei
enables investigation of the spin--dependent part of  the
amplitude of  scattering \cite{10,12,6b,antiprot}.

For low--energy neutrons, the phenomenon of "optical" spin
rotation (the phenomenon of nuclear precession of the neutron spin
in a nuclear pseudomagnetic field of a polarized target) was
predicted in \cite{4} and experimentally observed in \cite{5,6}.
This phenomenon is used  for measuring the spin--dependent forward
scattering amplitude of thermal neutrons
\cite{28,27,rins_7,rins_11,rins_8,rins_9}.

In contrast to neutrons, a charged particle moving in matter
undergoes Coulomb interaction with the atoms of matter, which
causes multiple scattering and rapid deceleration of the charged
particle due to ionization energy losses.

With decreasing  particle energy, the influence of Coulomb
interaction on particle scattering by the nucleus grows in
significance. In particular, when the energy of a positively
charged particle diminishes, the Coulomb repulsion suppresses
nuclear interaction between the incident particle and the target
nucleus and hence, the phenomenon of spin rotation due to nuclear
interaction. Conversely, a negatively charged particle
(antiproton, hyperon) is attracted to the nucleus and, as a
result, participates in nuclear interaction even at low energies.
As a consequence of this, spin rotation of a negatively charged
particle in polarized matter does not disappear at very low
energies either.

The present paper considers the influence of Coulomb interaction
on the phenomenon of "optical" spin rotation of negatively charged
particles moving in matter with polarized nuclei. It is shown that
because the density of the antiproton (negative hyperon) wave
function on the nucleus increases, the spin precession frequency
grows as the particle decelerates. As a result, spin rotation of
negatively charged particles becomes observable despite their
rapid deceleration. This provides information about the
spin--dependent part of  the scattering amplitude in the range of
low energies, where scattering experiments are practically
impossible to perform.

\section{Forward Scattering Amplitude of Negatively Charged Particles}

According to \cite{10,12,6b,4}, the spin rotation frequency
$\Omega_{\mathrm{nuc}}$ of a nonrelativistic particle passing
through a target with polarized nuclei can be expressed as
\begin{equation}
\label{a1} \Omega_{\mathrm{nuc}}=\frac{\Delta
\texttt{Re}U_{\mathrm{eff}}}{\hbar}=\frac{2\pi\hbar}{m}N \,P_t
\Delta \texttt{Re}f(0)\,,
\end{equation}
where $\Delta \texttt{Re}U_{\mathrm{eff}}$ is the difference
between the real parts of the effective potential energy of
interaction between the particle and the polarized target for
oppositely directed particle spins, $m$ is the mass of the
particle, $N$ is the number of nuclei in 1 cm$^3$, $P_t$ is the
degree of polarization of the target nuclei, $\Delta \texttt{Re}
f(0)$ is the difference between the real parts of the amplitudes
of coherent forward scattering for particles with oppositely
directed spins.

The scattering amplitude $f(0)$ is related to the T-matrix  as
follows (see, e.g. \cite{7,8}):
\begin{equation}
\label{a2}
f(0)=-\frac{m}{2\pi\hbar^2}\langle\Phi_a|T|\Phi_a\rangle\,,
\end{equation}
where $|\Phi_a\rangle$ is the wave function describing the initial
state of the system "incident particle--atom (nucleus)". The wave
function $|\Phi_a\rangle$ is the eigenfuction of the Hamiltonian
$\hat{H}_0=H_p(\vec{r}_p)+H_A(\vec{\xi}\,,
\vec{r}_{\mathrm{nuc}})$, i.e.,
$\hat{H}_0|\Phi_a\rangle=E_a|\Phi_a\rangle$; $H_p(\vec{r}_p)$ is
the Hamiltonian of the particle incident onto the target;
$\vec{r}_p$ is the particle coordinate; $H_A(\vec{\xi}\,,
\vec{r}_{\mathrm{nuc}})$ is the atomic Hamiltonian; $\vec{\xi}$ is
the set of coordinates of the atomic electron;
$\vec{r}_{\mathrm{nuc}}$ is the set of coordinates describing the
atomic nuclei.

The Hamiltonian $H$ describing the particle--nucleus  interaction
can be written as:
\begin{equation}
\label{a3} H=H_0+V_{\mathrm{Coul}} (\vec{r}_p\,,\vec{\xi}\,,
\vec{r}_{\mathrm{nuc}})+V_{\mathrm{nuc}}(\vec{r}_p\,,
\vec{r}_{\mathrm{nuc}})\,,
\end{equation}
where $V_{\mathrm{Coul}}$ is the energy of  Coulomb interaction
between the particle and the atom, $V_{\mathrm{nuc}}$ is the
energy of nuclear interaction between the particle and the atomic
nucleus.

According to the quantum theory of reactions \cite{7,8}, the
matrix element of the operator $T$ that describes the system
transition from the initial state $|\Phi_a\rangle$ into the final
state $|\Phi_b\rangle$ in this case of two interactions has the
form:
\begin{equation}
\label{a4} T_{ba}=\langle\Phi_b|V_{\mathrm{Coul}}
+V_{\mathrm{nuc}}|\Psi_a^+\rangle\,,
\end{equation}
where the wave function $\Psi_a^+$ satisfies the Schr\"{o}dinger
equation with the Hamiltonian of Eq. (\ref{a3}). At large
distances from the scatterer, the wave function $\Psi_a^+$ has the
asymptotics of the  diverging spherical wave \cite{7,8}.

The Shr\"{o}dinger equation  for the wave function $\Psi_a^+$ can
be written in the integral form:
\begin{equation}
\label{a5}
\Psi_a^+=\Phi_a+(E_a-H_0+i\varepsilon)^{-1}(V_{\mathrm{Coul}}
+V_{\mathrm{nuc}})\Psi_a^+\,.
\end{equation}
Let us introduce (see e.g. \cite{7,8,9} the wave functions
$\varphi_{a}^{(\pm)}$ describing the interaction between particles
and atoms via Coulomb interaction alone: ($V_{\mathrm{nuc}}=0$):
\begin{equation}
\label{a6} \varphi_{a}^{(\pm)}=\Phi_a+(E_a-H_0\pm
i\varepsilon)^{-1}V_{\mathrm{Coul}}\, \varphi_a^{(\pm)}\,,
\end{equation}
where the wave function $\varphi_a^{(-)}$ at large distances has
the asymtotics of a converging spherical wave \cite{7,8,9}.

Using Eq. (\ref{a6}), Eq. (\ref{a5}) for $\Psi_a^+$ can be written
in the form:
\begin{equation}
\label{a7} \Psi_a^+=\varphi_a^++(E_a-H_p-H_A-V_{\mathrm{Coul}}+
i\varepsilon)^{-1}V_{\mathrm{nuc}} \,\Psi_a^{+}\,.
\end{equation}
According to Eq. (\ref{a7}), the wave function $\Psi_a^+$ can be
represented as a sum of two waves: wave $\varphi_a^+$, defining
scattering due to the Coulomb interaction alone, and the wave that
appears as a result of scattering of wave $\varphi_a^+$ by the
nuclear potential $V_{\mathrm{nuc}}$ (the second term).

It eventually follows from Eqs. (\ref{a4}), (\ref{a7})  that the
matrix element $T_{ba}$ is representable as a sum of two terms
\cite{8}:
\begin{equation}
\label{a8}
T_{ba}=T_{ba}^{\mathrm{Coul}}+T_{ba}^{\mathrm{nuc\,Coul}}=
\langle\Phi_b|V_{\mathrm{Coul}}|\varphi_a^+\rangle+\langle\varphi_b^{(-)}|V_{\mathrm{nuc}}|\Psi_a^{(+)}\rangle\,.
\end{equation}
The first term $T_{ba}^{\mathrm{Coul}}$ describes the contribution
to the T-matrix that comes from the Coulomb scattering alone. The
second term describes the contribution to the T-matrix that comes
from nuclear scattering and takes account of the distortion of
waves incident onto the nucleus, which is caused by the Coulomb
interaction.

Equation (\ref{a8}) can also be presented in the form:
\begin{equation}
\label{a9}
T_{ba}=T_{ba}^{\mathrm{Coul}}+T_{ba}^{\mathrm{nuc\,Coul}}=
\langle\Phi_b|T_{\mathrm{Coul}}|\Phi_a\rangle+\langle\varphi_b^{(-)}|T_{\mathrm{nuc}}|\varphi_a^{(+)}\rangle\,,
\end{equation}
where the operator
\begin{equation}
\label{a10}
T_{\mathrm{Coul}}=V_{\mathrm{Coul}}+V_{\mathrm{Coul}}(E_a-H_0+i\varepsilon)^{-1}T_{\mathrm{Coul}}
\end{equation}
and the operator
\begin{eqnarray}
\label{a11}
T_{\mathrm{nuc}}&=&V_{\mathrm{nuc}}+V_{\mathrm{nuc}}(E_a-H_0- V_{\mathrm{Coul}}+i\varepsilon)^{-1}T_{\mathrm{nuc}}\nonumber\\
&=& V_{\mathrm{nuc}}+V_{\mathrm{nuc}}(E_a-H_0-
V_{\mathrm{Coul}}-V_{\mathrm{nuc}}+i\varepsilon)^{-1}V_{\mathrm{nuc}}\,.
\end{eqnarray}

Let us give a more detailed consideration of matrix element
$\langle\varphi_b^{(-)}|T_{\mathrm{nuc}}|\varphi_a^{(+)}\rangle$.
Because nuclear forces are short--range,  for this matrix element
the radius  of the domain of  integration is of the order of the
nuclear radius (of the order of the radius of action of nuclear
forces in the case of the proton). The Coulomb interaction
$V_{\mathrm{Coul}}$ in this domain is noticeably smaller than the
energy of nuclear interaction $V_{\mathrm{nuc}}$. We can therefore
neglect the Coulomb energy in the first approximation in the
denominator of Eq. (\ref{a11}), as compared to $V_{\mathrm{nuc}}$.

As a result, the operator $T_{\mathrm{nuc}}$ is reduced to the
operator describing a purely nuclear interaction between the
incident particle and the nucleus. The effect of Coulomb forces on
nuclear interaction is described by wave functions
$\varphi_{ba}^{(\pm)}$  differing from plane waves due to the
influence of Coulomb forces on the incident particle motion in the
area occupied by the nucleus  (distorted--wave approximation
\cite{8}).

In the range of antiproton energies of hundreds of
kiloelectronvolts and less, the de Broglie wavelength for
antiprotons scattered by the proton is larger than the radius of
action of nuclear forces (larger than the nuclear radius in the
case of antiproton scattering by the deuteron, $^3$He, etc.).
Therefore, in Eq. (\ref{a9}) for $T_{ba}^{\mathrm{nuc}}$, one can
remove the wave functions $\varphi_{a(b)}^{\pm)}$ outside the sign
of integration over the coordinate of the antiproton center of
mass, $\vec{R}_p$, at the location point of the nuclear center of
mass, $\vec{R}_{\mathrm{nuc}}$. As a result, one may write the
following relationship:
\begin{equation}
\label{a12}
T_{ba}^{\mathrm{nuc\,Coul}}=g_{ba}T_{ba}^{\mathrm{nuc}}=
\langle\varphi_b^{(-)}(\vec{R}_p=\vec{R}_{\mathrm{nuc}})|\varphi_a^{(+)}
(\vec{R}_p=\vec{R}_{\mathrm{nuc}})\rangle T_{ba}^{\mathrm{nuc}}\,,
\end{equation}
where $T_{ba}^{\mathrm{nuc}}$ is the matrix element describing a
purely nuclear interaction (in the absence of Coulomb interaction)
between the incident particle and the nucleus. The factor
$g_{ba}=\langle\varphi_b^{(-)}(\vec{R}_p=\vec{R}_{\mathrm{nuc}})|\varphi_a^{(+)}
(\vec{R}_p=\vec{R}_{\mathrm{nuc}})\rangle$ appearing in Eq.
(\ref{a12}) defines the probability to find the antiproton (the
negative hyperon, e.g. $\Omega^{-}\, , \Sigma^{-}$) at the point
of nucleus location.

From Eqs. (\ref{a3}), (\ref{a12}) follows the below expression for
the amplitude of coherent elastic zero--angle scattering:
\begin{equation}
\label{a13}
f(0)=-\frac{m}{2\pi\hbar^2}g_{aa}T_{aa}^{\mathrm{nuc}}=g_{aa}f_{\mathrm{nuc}}{(0)}\,,
\end{equation}
where $f_{\mathrm{nuc}}{(0)}$ is the amplitude of particle
scattering by the nucleus in the absence of Coulomb interaction,
$g_{aa}=\langle\varphi_b^{(-)}(\vec{R}_p=\vec{R}_{\mathrm{nuc}})|\varphi_a^{(+)}
(\vec{R}_p=\vec{R}_{\mathrm{nuc}})\rangle$ is the probability to
find the particle incident onto the nucleus at the point of
nucleus location.

Thus, the Coulomb interaction leads to the change in the value of
the amplitude of nuclear forward scattering. Let us estimate the
magnitude of this change.

According to \cite{9}, when a particle moves in the Coulomb field,
the probability $g_{aa}$ can be written in the form:
\begin{itemize}
\item for the case of repulsion, i.e., scattering of similarly
charged particles (e.g. scattering of  protons, deuterons by the
nucleus)
\begin{equation}
\label{a14}
g_{aa}^{\mathrm{rep}}=\frac{2\pi}{\kappa(e^{\frac{2\pi}{\kappa}}-1)}\,,\qquad
\kappa=\frac{v}{Z\alpha c}\,,
\end{equation}
where $v$ is the particle velocity, $Z$ is the charge of the
nucleus, $\alpha$ is the fine structure constant, $c$ is the speed
of light; \item for the case of attraction (e.g. scattering of
antiprotons, $\Omega^{-}\, , \Sigma^{-}$  by the nucleus)
\begin{equation}
\label{a15}
g_{aa}^{\mathrm{att}}=\frac{2\pi}{\kappa(1-e^{-\frac{2\pi}{\kappa}})}\,,
\end{equation}
\end{itemize}

With decreasing particle energy (velocity), $\kappa$ diminishes,
and for such values of $\kappa$ when $\frac{2\pi}{\kappa}\geq1$,
one can write
\begin{equation}
\label{a16} g_{aa}^{\mathrm{rep}}=\frac{2\pi\alpha Z
c}{v}\,e^{-\frac{2\pi\alpha Z c}{v}}\,,
\end{equation}
\begin{equation}
\label{a17} g_{aa}^{\mathrm{att}}=\frac{2\pi\alpha Z c}{v}\,.
\end{equation}

According to Eq. (\ref{a16}), with decreasing particle (proton,
deuteron) energy, the amplitude $f(0)$ diminishes rapidly because
of repulsion. For negatively charged particles (antiprotons,
$\Omega^{-}\, , \Sigma^{-}$-hyperons, and so on), the amplitude of
coherent elastic zero--angle scattering grows with decreasing
particle energy (velocity).

These results for the amplitude $f(0)$ generalize a similar,
well-known  relationship for taking account of the Coulomb
interaction effect on the cross section of inelastic processes,
$\sigma_r$, \cite{9}.

So in the range of low energies, the amplitude of coherent elastic
forward scattering of the antiproton (negative hyperon) by the
nucleus can be presented in the form:
\begin{equation}
\label{a18} f(0)=\frac{2\pi \alpha Z c}{v}
f_{\mathrm{nuc}}=\frac{2\pi\alpha Z c}{v}\texttt{Re}\,
f_{\mathrm{nuc}}+\frac{i}{2}\frac{\alpha
Z}{\lambda_c}\sigma_{\mathrm{tot}}\,,
\end{equation}
where $\lambda_c=\frac{\hbar}{mc}$, $\sigma_{\mathrm{tot}}$ is the
total cross section of nuclear interaction between the particle
and the scatterer. In deriving Eq. (\ref{a18}), the optical
theorem ${\texttt{Im}\, f(0)=\frac{k}{4\pi}\sigma_{\mathrm{tot}}}$
was applied, where $k$ is the wave number of the incident
particle.

\section{Effective Potential Energy of Negatively Charged
Particles in Matter}

Using the amplitude $f(0)$, one can write the expression for the
refractive index $\hat{n}$ of a spin particle in matter with
polarized nuclei, as well as the expression for the effective
potential energy $\hat{U}_{\mathrm{eff}}$ of interaction between
this particle and matter \cite{10,6b}:
\begin{equation}
\label{a19} \hat{n}^2=1+\frac{4\pi\,
N}{k^2}\hat{f}(0)\qquad\mbox{and}\qquad
\hat{U}_{\mathrm{eff}}=-\frac{2\pi\hbar^2}{m} N\hat{f}(0)\,,
\end{equation}
where in the case considered here, $\hat{f}(0)=\frac{2\pi\alpha Z
c}{v}\hat{f}_{\mathrm{nuc}}(0)$ is the amplitude of coherent
elastic zero-angle scattering being the operator in the particle
spin space.

The amplitude $\hat{f}(0)$ depends on the vector polarization
$\vec{P}_t$ of the target nuclei and can be presented in the form:
\begin{equation}
\label{a20} \hat{f}(0)=A_0+A_1(\hat{\vec{S}}\, \vec{P}_t)+
A_2(\hat{\vec{S}}\, \vec{e})(\vec{e}\, \vec{P}_t)\,,
\end{equation}
where $A_0$ is the scattering amplitude independent  of the
incident particle spin, $\hat{\vec{S}}$ is the particle spin
operator, $\vec{e}$ is the unit vector in the direction of the
particle momentum. If the spin of the target nuclei $I\geq1$, then
the addition depending on the target tensor polarization also
appears \cite{10,6b}.

Correspondingly, the effective potential energy
$\hat{U}_{\mathrm{eff}}$ of particle interaction with polarized
matter looks like
\begin{equation}
\label{a21} \hat{U}_{\mathrm{eff}}=-\frac{2\pi\hbar^2}{m}
N(A_0+A_1(\hat{\vec{S}}\, \vec{P}_t)+ A_2(\hat{\vec{S}}\,
\vec{e})(\vec{e}\, \vec{P}_t))\,.
\end{equation}
Expression (\ref{a21}) can be written as
\begin{equation}
\label{a22} \hat{U}_{\mathrm{eff}}={U}_{\mathrm{eff}}+
\hat{V}_{\mathrm{eff}} (\vec{P}_t)\,,
\end{equation}
where
\begin{equation}
\label{a23} {U}_{\mathrm{eff}}=-\frac{2\pi\hbar^2}{m} N A_0\,,
\end{equation}
\begin{equation}
\label{a24}
\hat{V}_{\mathrm{eff}}(\vec{P}_t)=-\vec{\mu}\vec{G}=-\frac{\mu}{S}(\hat{\vec{S}}\,
\vec{G})\,,
\end{equation}
where $\mu$ is the particle magnetic moment,
\begin{equation}
\label{a25} \vec{G}=\frac{2\pi\hbar^2 S}{m\mu}
N(A_1\vec{P}_t+A_2\vec{e}(\vec{e}\, \vec{P}_t))\,,
\end{equation}

Recall now that the energy of interaction between the magnetic
moment $\mu$ and a magnetic field $\vec{B}$ is as follows:
\begin{equation}
\label{a26} V_{\mathrm{mag}}= -(\vec{\mu}\,
\vec{B})=-\frac{\mu}{S}(\hat{\vec{S}}\vec{B})\,.
\end{equation}
Expressions (\ref{a24}) and (\ref{a25}) are identical. Therefore,
$\vec{G}$ can be interpreted as the effective pseudomagnetic field
acting on the spin of the particle moving in matter with polarized
nuclei and appears due to nuclear interaction between the incident
particles and the scatterers. Similarly to particle spin
precession in an external magnetic field, particle spin precesses
in field $\vec{G}$. This phenomenon was called the nuclear
precession of the particle spin, first described for slow neutrons
in \cite{4} and then observed in \cite{5,6}.

It is worth noting that the amplitudes $A_1$ and $A_2$ determining
the effective field $\vec{G}$ depend on the energy, the
orientation of vectors $\vec{e}$ and $\vec{P}_t$ and are complex
values. As a result, unlike the magnetic field $\vec{B}$, the
effective pseudomagnetic field $\vec{G}$ depends on the particle
energy and the orientation of vectors $\vec{e}$ and $\vec{P}_t$.
The field $\vec{G}$ is a complex value. The spin precession
frequency in this field is determined by  $\texttt{Re} \vec{G}$.

From Eq. (\ref{a18}) follows that in the range of low energies,
$\hat{U}_{\mathrm{eff}}$ can be presented in the form:
\begin{equation}
\label{a27} \hat{U}_{\mathrm{eff}}= \frac{2\pi \alpha Z
c}{v}\hat{U}_{\mathrm{eff}}^{\mathrm{nuc}}\,,
\end{equation}
where $\hat{U}_{\mathrm{eff}}^{\mathrm{nuc}}$ coincides in form
with that in Eq. (\ref{a21}) with the amplitudes $A_0$, $A_1$, and
$A_2$ replaced by $A_0^{\mathrm{nuc}}$, $A_1^{\mathrm{nuc}}$, and
$A_2^{\mathrm{nuc}}$ calculated ignoring the Coulomb interaction.

According to Eq. (\ref{a27}), with decreasing particle velocity,
$\hat{U}_{\mathrm{eff}}$ grows, as well as the field $\vec{G}$,
and particle spin precession in this field:
\begin{equation}
\label{a28} \Omega_{\mathrm{pr}}\sim
\texttt{Re}\,G\sim\frac{1}{v}\,.
\end{equation}
Let us take a somewhat different view of this issue.

In view of Eqs. (\ref{a21}), (\ref{a24}), and (\ref{a27}),
$U_{\mathrm{eff}}$ depends on the orientation of vectors $\vec{e}$
and $\vec{P}_t$.

According to Eq. (\ref{a21}), two simpler cases can be
distinguished:

Let $\vec{e}\perp \vec{P}_t$, then
\begin{equation}
\label{a29}
\hat{U}_{\mathrm{eff}}^{\perp}=-\frac{2\pi\hbar^2}{m}\,
N(A_0+A_1(\hat{\vec{S}} \vec{P}_t))\,,
\end{equation}

If the target polarization vector $\vec{P}_t$ is directed along
vector $\vec{e}$, then
\begin{equation}
\label{a30}
\hat{U}_{\mathrm{eff}}^{\parallel}=-\frac{2\pi\hbar^2}{m}\,
N(A_0+(A_1+A_2)(\hat{\vec{S}}\vec{P}_t))\,,
\end{equation}

Direct the quantization axis parallel to the polarization vector
$\vec{P}_t$. Hence for particle states with magnetic quantum
number $M_s$, one can write the below expression for
$U_{\mathrm{eff}}^{\perp}(M_s)$, which follows from Eq.
(\ref{a29}):
\begin{equation}
\label{a31} U_{\mathrm{eff}}^{\perp}(M_s)=-\frac{2\pi\hbar^2}{m}\,
N(A_0+A_1M_s P_t)\,.
\end{equation}
$U_{\mathrm{eff}}^{\parallel}$ is obtained by replacing $A_1$ with
$A_1+A_2$.

In the case of antiprotons ($M_s=\pm\frac{1}{2}$), from Eq.
(\ref{a22}) one can obtain two values of the effective potential
energy, depending on the antiproton spin orientation:
\begin{equation}
\label{a32}
U_{\mathrm{eff}}^{\perp}=-\frac{2\pi\hbar^2}{m}N(A_0\pm\frac{1}{2}A_1P_t)\,,
\end{equation}
\begin{equation}
\label{a33}
U_{\mathrm{eff}}^{\parallel}=-\frac{2\pi\hbar^2}{m}N(A_0\pm\frac{1}{2}(A_1+A_2)P_t)\,,
\end{equation}

The difference of the real parts of these energies defines the
spin precession frequency of the antiproton in matter with
polarized nuclei:
\begin{eqnarray}
\label{a34}
\Omega^{\perp}_{\mathrm{pr}}& = &\frac{\texttt{Re}(U_{\mathrm{eff}}^{\perp}(+\frac{1}{2})-U^{\perp}_{\mathrm{eff}}(-\frac{1}{2}))}{\hbar}\\
\nonumber & = &-\frac{2\pi\hbar}{m}\,N P_t \texttt{Re}\,
A_1=-\frac{2\pi\hbar}{m}\cdot\frac{2\pi \alpha Z c}{v} N P_t\,
\texttt{Re}\, A_{1\,\mathrm{nuc}}\,,\\
\Omega_{\mathrm{pr}}^{\parallel}&=&-\frac{2\pi\hbar}{m}\cdot\frac{2\pi\alpha
Z c}{v} NP_t\,\texttt{Re}(A_{1\mathrm{nuc}}+A_{2\,\mathrm{nuc}})
\end{eqnarray}
Thus, the spin precession frequency of a negatively charged
particle grows with decreasing energy (velocity):
\[
\Omega_{\mathrm{pr}}\sim \frac{1}{v}\,.
\]

\section{Spin Rotation Angle of Low--Energy Antiprotons in
Polarized Matter}

Let us estimate the magnitude of the effect. Because in the range
of low energies the de Brogile wavelength of a particle is much
larger than the nuclear radius, in making estimations we shall
concentrate on $S$-scattering (one should bear in mind, though,
that the analysis of the antiproton--proton interaction in
protonium has shown that at low energies,  $P$-waves also
contribute to antiproton--proton interaction \cite{1}). In
$S$-wave scattering, the amplitude $A_2^{\mathrm{nuc}}$ equals
zero.

The amplitudes $A_0^{\mathrm{nuc}}$ and $A_1^{\mathrm{nuc}}$ can
be expressed in terms of the amplitudes $a^+$ and $a^-$, where
$a^+$ is the scattering amplitude in the state with total momentum
$I+\frac{1}{2}$, and $a^-$ is the same in the state with total
momentum $I-\frac{1}{2}$ ($I$ is the nuclear spin) \cite{9}:
\begin{eqnarray}
\label{a36}
A_0^{\mathrm{nuc}}&=&\frac{I+1}{2I+1}a^++\frac{I}{2I+1}
a^{-}\\
A_1^{\mathrm{nuc}}&=&\frac{2I}{2I+1}(a^+-a^-)\,.\nonumber
\end{eqnarray}
As a consequence, one can write the following expression for
$\Omega_{\mathrm{pr}}$:
\begin{equation}
\label{a37}
\Omega_{\mathrm{pr}}=-\frac{2\pi\hbar}{m}\cdot\frac{2\pi \alpha Z
c}{v} \, NP_t\frac{2I}{2I+1} \,\texttt{Re}(a^+-a^-)\,.
\end{equation}

When antiprotons pass through a hydrogen target ($I=\frac{1}{2}$),
we have:
\begin{equation}
\label{a38} \Omega_{\mathrm{pr}}=\frac{\pi\hbar}{m}\cdot\frac{2\pi
\alpha  c}{v} \, NP_t\,\texttt{Re}(a^+-a^-)\,.
\end{equation}
The factor $\frac{2\pi\alpha c}{v}$ makes Eq. (\ref{a38})
different from the equation for spin precession frequency of slow
neutrons moving in a target with polarized protons.

Recall that in the range of low energies, the amplitudes $a^+$ and
$a^-$ are often expressed in terms of the scattering lengths $b^+$
and $b^-$ \cite{9}:
\[
a^+=-b^+\qquad \mbox{and}\qquad a^-=-b^-\,.
\]

When the neutron is scattered by the proton, $b^+=5.39\cdot
10^{-13}$\,cm, $b^-=-2.37\cdot 10^{-12}$\,cm \cite{9}. As a
result, in the case of $n-p$ scattering, for the amplitude
$A_0^{\mathrm{nuc}}$ (see Eq. (\ref{a36})) we have $A_0\approx
-1.9\cdot 10^{-13}$\, cm, while for the amplitude
$A_1^{\mathrm{nuc}}$, we have $A_1^{\mathrm{nuc}}\approx 1.46\cdot
10^{-12}$. As is seen, $A_1\gg |A_0|$.

In the case of antiproton--proton interaction, the
spin--independent part of the scattering length, "$b$", is $b\sim
10^{-13}$\,cm \cite{1}, which is comparable with
$|A_0^{\mathrm{nuc}}|$ for the case  of $n-p$ scattering. For
antiproton scattering by the proton, the magnitude of $A_1$ is
unknown in the considered range of low energies.

To estimate the magnitude of $A_1$, let us assume that
$\texttt{Re}\,A_1^{\mathrm{nuc}}$ in the case of $\bar{p}-p$
scattering is comparable with $\texttt{Re}\,A_1^{\mathrm{nuc}}$ in
the case of $n-p$ scattering, i.e.,
$\texttt{Re}\,A_1^{\mathrm{nuc}}$ is of the order of $10^{-12}$.

We can finally obtain the following estimation for the antiproton
spin precession frequency in matter with polarized protons:
\begin{equation}
\label{a39} \Omega_{\mathrm{pr}}=\frac{2\pi\alpha
c}{v}\cdot\frac{2\pi\hbar}{m} NP_t
\texttt{Re}\,A_1\simeq\frac{2\pi\alpha c}{v} 6 \cdot 10^7
\frac{N}{N_{\mathrm{l}}} P_t\,,
\end{equation}
where $N_l$ is the number of atoms in 1 cm$^3$ of liquid hydrogen,
$N_{l}\simeq 4.25\cdot 10^{22}$. It will be recalled that
$\frac{2\pi \alpha c}{v}\gg 1$ for slow antiprotons with velocity
${v< 10^9}$\,cm/s.

Let us estimate now the spin rotation angle of the antiproton.
Eventually, the antiproton that moves in matter soon appears
captured by the proton and forms a bound state, protonium. We
shall therefore estimate the magnitude of the spin rotation angle
during the characteristic time $\tau$ necessary for the antiproton
to be captured into a bound state:
\[
\tau\sim\frac{1}{N v\sigma_{\mathrm{pr}}}\,
\]
where $\sigma_{\mathrm{pr}}$ is the cross section of the protonium
formation.

The antiproton spin rotation angle $\vartheta$  during this time
can be estimated using formula:
\begin{equation}
\label{a40}
\vartheta\sim\Omega_{\mathrm{pr}}\tau=\frac{2\pi\hbar}{m}\cdot\frac{2\pi\alpha
c}{v^2}\, P_t\,
\frac{\texttt{Re}A_1}{\sigma_{\mathrm{pr}}}=\frac{2\pi^2\hbar}{E}\alpha
c\frac{\texttt{Re}A_1}{\sigma_{\mathrm{pr}} }P_t\,,
\end{equation}
where $E=\frac{mv^2}{2}$ is the antiproton energy.

 Antiproton beams with the energy of hundreds of electronvolts and smaller are presently
 available. According to \cite{11}, in the range of energies  higher than 10\,e\,V, the cross section  of
 antiproton capture by hydrogen with the formation of protonium is $\sigma_{\mathrm{pr}}\leq
 10^{-18}$.  As a result, the spin rotation angle for antiprotons with the energy of $10^2$\, eV can be
 estimated as $\vartheta\simeq 6\cdot10^{-2} P_t$.
 For antiproton energies of 20\,eV, the same is $\vartheta\sim 3\cdot 10^{-1}
 P_t$.
When antiprotons are decelerated in a polarized gaseous target,
the degree of proton polarization is close to unity. As a
consequence, the rotation angle in such a target reaches quite
appreciable values, $\vartheta\sim 10^{-1}$, giving hope for
experimental observation of the effect. Thus, this effect  will be
applicable for $\texttt{Re}\,A_1$ measurement in the region where
the scattering experiments are really difficult to realize.  Note
that for studying the polarization state, one can use a polarized
target with nuclei having a large $Z$ (in fact, a thin film, in
which case nuclei in a static magnetic field can be polarized far
more readily).

After the antiproton has been captured into a bound state by
hydrogen, a new atom appears, called protonium. This atom is
neutral and, moving in matter, undergoes interactions with a
nuclear pseudomagnetic field of polarized matter in a similar way
as a neutron.  Such interaction leads to splitting and shifts of
energy levels of excited and ground states of the protonium (and
similar atoms, such as $\bar{p}$\,d, $\bar{p}^3$He), as well as to
spin rotation and oscillations of the excited and ground states of
these atoms.

In this regard, we would like to specially mention the system
$\bar{p}^3$He for having, like $\bar{p}^4$He, a considerably long
lifetime ($\approx 10^{-6}$\,s).
The long lifetime of the system $\bar{p}$He  makes it possible to
use presently available laser-microwave techniques
\cite{wid,yam,hay} for studying the systems's  values of
splitting, shifts, and  widths of energy levels caused by the
nuclear pseudomagnetic field produced by the gas from polarized
atoms in which the system $\bar{p}$He moves
 (e.g. $\bar{p}$ capture in a mixture of
polarized H, D, $^3$He gases  or  in polarized H, D, or $^3$He
mixed with $^4He$).
Because the nuclear pseudomagnetic field depends on the spin part
of the scattering amplitude, by measuring the shift and width of
the level one can obtain information about the amplitude of
antiproton scattering by  atomic nuclei  of the gas in which the
system $\bar{p}$He ($\bar{p}^3$He, $\bar{p}^4$He) moves.
In such experimental arrangement, the system $\bar{p}$He acts as a
probe. Analyzing how the shifts, splitting and widths of the
energy levels depend on, e.g., the density of polarized gas, one
can obtain the  values of the shift (splitting) and width of the
level that we are concerned with \cite{6b} and thus obtain
information about the amplitude of $\bar{p}$ scattering by the
nuclei of neighboring atoms.
Note here that in the considered experimental arrangement, the
antiproton beam incident onto the target can be unpolarized.

It is worth noting in conclusion that, as we have shown in
\cite{oscil_3,rins_20,rins_21}, in annihilation of positronium
(the bound state of positronium $e^+$ and electron $e^-$), one can
observe quantum time--oscillations of the counting rate of the
annihilation products for ortho-positronium placed in a magnetic
field. These quantum oscillations were experimentally observed and
laid the basis for the methods of studying fundamental
interactions and the properties of matter \cite{6b}.

Similar phenomena of quantum oscillations of the counting rate of
the annihilation products, depending on the time passed from the
moment when the particle entered the target, can also be observed
for long--lived systems ($\bar{p}^3$He, $\bar{p}^4$He). It should
be noted  that quantum oscillations of the counting rate of the
annihilation products occur at the frequencies determined by the
difference of energies between the levels of the bound system. For
this reason,  use of the external magnetic field  enables changing
distances between the  energy levels of the considered atoms and
thus selecting optimal conditions for observation of the
phenomenon of quantum oscillations.

\bigskip

The author is grateful to K. Fregat for valuable discussions.

\end{document}